\documentclass[amsmath,amssymb,twocolumn,floatfix, nofootinbib]{revtex4-2}

\usepackage{graphicx}
\usepackage{dcolumn}
\usepackage{bm}
\usepackage{hyperref}

\providecommand{\mislash}[1]{#1 \mspace{-10.0mu} \slash}

\providecommand{\proarrow}[0]{\rightarrow}

\providecommand{\dif}[0]{\mathrm{d}}

\providecommand{\proname}[2]{#1 \proarrow #2}

\providecommand{\miim}[1]{{\rm Im} \left[ #1 \right]}

\providecommand{\order}[1]{{\cal O} \left( #1 \right)}

\providecommand{\g}[2]{\gamma^{#1}_{#2}}

\providecommand{\gp}[2]{\gamma^{#1 \; '}_{#2}}

\providecommand{\mhdos}[0]{m_{\eta}}
\providecommand{\g}[2]{\gamma^{#1}_{#2}}

\providecommand{\gp}[2]{\gamma^{#1 \; '}_{#2}}

\begin{document}

\title{Low-scale leptogenesis in the scotogenic model: spectator processes and benchmark points}

\author{J.~Racker}
\email{jracker@unc.edu.ar}
\affiliation{Instituto de Astronom\'{\i}a Te\'orica y Experimental (IATE), Consejo Nacional de Investigaciones Cient\'{\i}ficas y T\'ecnicas
(CONICET)~- Universidad Nacional
de C\'ordoba (UNC), Laprida 854, X5000BGR, Córdoba, Argentina
}
\affiliation{
 Observatorio Astron\'omico de C\'ordoba (OAC), Universidad Nacional de C\'ordoba (UNC), Laprida 854, X5000BGR, Córdoba, Argentina
}

\begin{abstract}
We study leptogenesis from the decay of the lightest sterile neutrino in the scotogenic model with a scalar dark matter candidate. Our analysis focuses on the possible exponential suppression of washouts for sizable values of the inert Higgs mass and the crucial role of some spectator processes for this to happen. We show that leptogenesis can be successful for TeV-scale masses of the lightest sterile neutrino for both, thermal and zero, initial abundances of the neutrinos. Moreover, leptogenesis is viable for Yukawa couplings of the heavier sterile neutrinos large enough to yield observable charged lepton flavor violation processes in current and planned experiments. 
\end{abstract}

\maketitle

\section{Introduction}

The baryon density of the Universe has been determined to better than one percent by {\it Planck}~\cite{Planck:2018vyg} and is consistent with the Big-Bang nucleosynthesis prediction of light element abundances (see e.g.~the recent analysis in~\cite{Yeh:2022heq}). Given that the Universe does not seem to contain significant amounts of antimatter~\cite{Cohen:1997ac}, the baryon density is also a measure of the baryon asymmetry of the Universe (BAU), which, normalized to the entropy density, $s$, is equal to $Y_B\equiv n_B/s \simeq 8.7 \times 10^{-11}$~\cite{Planck:2018vyg}.
The BAU is one of the strongest evidences of physics beyond the Standard Model (SM) because, within the SM, there is too little CP violation~\cite{Gavela:1993ts, Gavela:1994dt} and not enough departure from equilibrium (see e.g.~the review~\cite{Bodeker:2020ghk}), these being two of the three Sakharov's conditions~\cite{Sakharov:1967dj} to generate dynamically a baryon asymmetry.

Currently, baryogenesis via leptogenesis in neutrino mass models~\cite{Fukugita:1986hr}, notably in the type I seesaw, is one of the best motivated explanations for the BAU, as baryogenesis occurs naturally in simple extensions of the SM which were not even proposed to address this cosmological puzzle in the first place. The baryon asymmetry is the result of electroweak sphalerons~\cite{Kuzmin:1985mm} reprocessing part of the lepton asymmetry generated in the production and/or decay of some exotic particle of mass $M$ related to the origin of neutrino masses. 
Unfortunately, these scenarios typically require masses too high to allow for experimental exploration in foreseeable experiments (beyond the connection with light neutrino masses). In the type I seesaw with hierarchical heavy neutrino masses, the Davidson-Ibarra bound~\cite{Davidson:2002qv} for the CP asymmetry in neutrino decays, which is proportional to the masses of the heavy and light neutrinos (relative to the electroweak scale), implies that $M \gtrsim 10^8-10^9$~GeV. More generally, low-scale leptogenesis generally suffers from strong washout processes that erase the asymmetry because the Hubble rate ($H$), which is the source of departure from equilibrium, gets milder as the temperature ($T$) of the leptogenesis epoch decreases. Specifically, $H \propto T^2/m_{\rm Pl}$, and therefore the Planck mass ($m_{\rm Pl}$) sets a scale for departures from equilibrium, which occur when the rates of some relevant processes become smaller than $H$.

This has motivated many works on model building with the main purpose of increasing testability. 
However it has also been realized that low-scale leptogenesis can be successful in particular regions of the parameter space of models as simple and well motivated as the type I seesaw. In this model, it is possible to have leptogenesis via neutrino oscillations~\cite{Akhmedov:1998qx, Asaka:2005pn} during the production of sterile neutrinos heavier than $\sim 100$~MeV, 
as well as resonant  leptogenesis~\cite{Pilaftsis:2003gt} during the decay of almost-degenerate neutrinos with masses larger than $\order{10^2}$~GeV. Indeed the two mechanisms are related~\cite{Klaric:2020phc, Jukkala:2021sku, Racker:2022ibr} and the allowed regions of parameter space overlap~\cite{Klaric:2020phc, Klaric:2021cpi} (see also~\cite{Hambye:2016sby}). Moreover, it is  important to stress that some regions of parameter space are compatible with leptogenesis and measurable active-sterile mixings (see~\cite{Drewes:2017zyw} for a review and references).

Remarkably, the type I seesaw with one KeV-scale neutrino may also provide a dark matter (DM) candidate, however the correct amount of BAU and DM can only be obtained in outstandingly fine tuned regions of the parameter space~\cite{Asaka:2005pn, Canetti:2012kh, Ghiglieri:2019kbw, Ghiglieri:2020ulj}. The phenomenology changes substantially if, together with the sterile neutrinos, a second Higgs doublet is added to the SM particle content, and the new particles are required to be odd under a discrete $Z_2$ symmetry (while the SM fields are even). This is the scotogenic model~\cite{Ma:2006km}, which is arguably the simplest and most studied model to explain and relate the smallness of neutrino masses with DM. The $Z_2$ symmetry implies that there are no tree-level contribution to neutrino masses, which are instead generated radiatively, and that the lightest odd particle is stable, becoming a DM candidate if it is one of the electrically neutral species. 

Soon after its proposal it was realized that baryogenesis via leptogenesis is also possible in this model and that the lower bound on $M$ due to the Davidson-Ibarra bound could be relaxed~\cite{Ma:2006fn, Hambye:2009pw}. However, only estimates were presented for the crucial washout effects. Indeed more detailed treatments including all neutrino oscillation data were performed in~\cite{Kashiwase:2012xd, Kashiwase:2013uy}, concluding that successful leptogenesis with TeV-scale masses seemed possible only for quasi-degenerate neutrinos resorting to the resonant leptogenesis mechanism. A more exhaustive exploration of the  parameter space for leptogenesis in the scotogenic model was performed in~\cite{Racker:2013lua} and~\cite{Hugle:2018qbw}. The results of~\cite{Racker:2013lua} suggested that non-resonant leptogenesis at the TeV-scale from the decay of the lightest sterile neutrino was feasible, either by assuming an initial thermal abundance for this species and/or for sizable masses of the inert Higgs (the new Higgs doublet) in its role of being one of the decay products of the sterile neutrinos. However, the work~\cite{Racker:2013lua} centered on the mechanisms to avoid the strong washouts and it did not attempt to fit neutrino oscillation data. Moreover, it performed a simplified treatment of some crucial spectator processes (i.e., processes which, while not being the source of the lepton asymmetry, they are active during leptogenesis and influence the resultant BAU). Instead, in~\cite{Hugle:2018qbw} the exploration of the parameter space was performed accounting for all neutrino oscillation data (via a Casas-Ibarra~\cite{Casas:2001sr} parametrization of the Yukawa couplings), but the flavor structure in the transport equations~\cite{Barbieri:1999ma, Endoh:2003mz, Pilaftsis:2004xx, Pilaftsis:2005rv, Abada:2006fw, Nardi:2006fx, Abada:2006ea, Blanchet:2006ch} was neglected and, more importantly, the effects of the inert Higgs mass on leptogenesis were not considered. The lower bound on the mass of the lightest sterile neutrino, assuming a thermal initial abundance, was found to be around 10~TeV~\footnote{Other studies of leptogenesis in the scotogenic model include~\cite{Baumholzer:2018sfb} (for leptogenesis via neutrino oscillations), \cite{Borah:2018rca} (where the inert Higgs doublet is coupled non-minimally to gravity to achieve inflation), \cite{Mahanta:2019gfe} (for $N_2$-leptogenesis with fermionic DM), \cite{Clarke:2015hta} (for natural leptogenesis in two Higgs doublet models), and see also the review and references in~\cite{Cai:2017jrq}.}.

Here we use an extended set of flavored Boltzmann Equations (BEs), including explicitly some crucial spectator processes, to show that for sizable values of the inert Higgs mass, the parameter space compatible with DM, neutrino masses and successful leptogenesis opens up considerably. In particular we will show that the BAU can be obtained with TeV-scale masses for both, thermal and zero, initial abundances of the sterile neutrinos and that  leptogenesis is viable for Yukawa couplings large enough to yield observable charged lepton flavor violation (CLFV) processes in current and planned experiments. The detailed treatment of the exponential Boltzmann suppression of washouts due to a massive particle different from its antiparticle is interesting by itself and may be applied to other extensions of the SM (the less trivial part being that the asymmetry that builds up in the massive field also induces washouts).

\section{The model and Boltzmann equations} 
The scotogenic model extends the SM by a second Higgs doublet $\eta$ and (two or) three singlet fermions $N_i$ (aka sterile neutrinos). The model also has a discrete $Z_2$ symmetry under which the new fields are odd and the SM fields are even. The Lagrangian can be expressed as
\begin{eqnarray*}
\mathcal{L} &=& \mathcal{L}_{\text{SM}} + \frac{i}{2} \overline{N}_i 
  \mislash{\partial} N_i + (D_\mu \eta)^\dag (D^\mu \eta) - \frac{M_i}{2} \overline{N}_i N_i \nonumber \\ && - V(\phi,\eta) - (h_{\alpha i}\, \overline{N_i}\, {\widetilde \eta}^\dag\, \ell_\alpha 
+ {\rm h.c.})\, ,
\label{eq:lagrangian}
\end{eqnarray*}
where $\mathcal{L}_{\text{SM}}$ is the SM Lagrangian (without the scalar potential, which is included below), $\alpha=e, \mu, \tau$ and 
$i=1,2, 3$ are flavor indices, $\ell_\alpha$ are the leptonic $SU(2)$ doublets, 
$\eta=(\eta^+,(\eta_R+i\eta_I)/\sqrt{2})^T$ is the inert Higgs ($\widetilde \eta =i\sigma_2 \eta^*$), $\phi$ is the SM Higgs and the scalar potential $V(\phi, \eta)$ can be written in terms of real parameters $m_\phi^2, m_\eta^2, \lambda_{1,2,3,4,5}$ as
\begin{eqnarray*}
   &&V(\phi, \eta) = m_\phi^2 \phi^\dag \phi + m_\eta^2 \eta^\dag \eta + \frac{\lambda_1}{2}(\phi^\dag \phi)^2 + \frac{\lambda_2}{2} (\eta^\dag \eta)^2 \nonumber \\ &&+ \lambda_3 (\phi^\dag \phi)(\eta^\dag \eta) + \lambda_4 (\eta^\dag \phi)(\phi^\dag \eta) + \frac{\lambda_5}{2}\left[(\eta^\dag \phi)^2 + (\phi^\dag \eta)^2\right]. \nonumber
\end{eqnarray*}
After electroweak symmetry breaking, the new physical scalar states have masses
\begin{eqnarray}
    m_{\eta^\pm}^2 &=& m_\eta^2 + \lambda_3 v^2, \nonumber \\
     m_{\eta_R}^2 &=& m_\eta^2 + (\lambda_3 + \lambda_4 + \lambda_5) v^2, \nonumber \\
     m_{\eta_I}^2 &=& m_\eta^2 + (\lambda_3 + \lambda_4 - \lambda_5) v^2, 
\end{eqnarray}
with $v=246/\sqrt{2}$~GeV the vacuum expectation value of the Higgs. 

Here we will consider the case that one of the neutral components of $\eta$ is the lightest odd particle under the $Z_2$ symmetry.
For definiteness we take $\lambda_5 > 0$ so that $\eta_I$ is the DM candidate and restrict our study to the mass window $530~{\rm GeV} \lesssim m_{\eta_I} \lesssim 20~{\rm TeV}$, were the correct relic abundance can be obtained with a proper choice of the parameters (for bounds and related studies see e.g.~\cite{Barbieri:2006dq, LopezHonorez:2006gr, Hambye:2009pw, LopezHonorez:2010eeh, Dolle:2009fn,  LopezHonorez:2010tb, Goudelis:2013uca, Krawczyk:2013jta, Klasen:2013jpa, Arhrib:2013ela, Ilnicka:2015jba, Garcia-Cely:2015khw, Diaz:2015pyv,  Belyaev:2016lok, Borah:2017dfn, Eiteneuer:2017hoh, Avila:2021mwg}).

Also due to the $Z_2$ symmetry there are no tree-level contributions to the light neutrino masses, which are induced at one-loop order and therefore they can be naturally small for lower values of $M_i$ compared to the type I seesaw. Moreover, neutrino masses are also suppressed by $\lambda_5$, which is a natural parameter in the sense of ’t Hooft, as a conserved lepton number can be defined in the limit $\lambda_5 \to 0$. At one-loop order the mass matrix of the active neutrinos is given by~\cite{Ma:2006km,Merle:2015ica}
\begin{equation}
\label{eq:mnu}
(m_\nu)_{\alpha \beta}=(h\, \Lambda\, h^T)_{\alpha \beta},
\end{equation}
where $\Lambda$ is a diagonal matrix with entries
\begin{equation}
\label{eq:lambda}
    \Lambda_{ii}=\frac{M_i}{32 \pi^2} \left[\frac{m_{\eta_R}^2}{m_{\eta_R}^2-M_i^2}\ln{\frac{m_{\eta_R}^2}{M_i^2}} - \frac{m_{\eta_I}^2}{m_{\eta_I}^2-M_i^2}\ln{\frac{m_{\eta_I}^2}{M_i^2}}\right].
\end{equation}
The mass matrix $m_\nu$ can be diagonalized by the Pontecorvo-Maki-Nakagawa-Sakata (PMNS) mixing matrix, $U$, to yield the light neutrino masses, $U^T \,m_\nu\, U= d_\nu =\rm{diag}(m_1,m_2,m_3)$. From these relations it is possible to obtain a Casas-Ibarra~\cite{Casas:2001sr} parametrization for the Yukawa coupling matrix $h=(h_{\alpha i})$ in terms of a complex orthogonal matrix $R$,
\begin{equation}
\label{eq:ci}
    h=U^*\, \sqrt{d_\nu} \, R^T\, \sqrt{\Lambda}^{\,-1}.
\end{equation}
In turn, the matrix $R$ can be written as the product of three rotations by complex angles $z_{23}, z_{13}$ and $z_{12}$, $R=R_{23}\,R_{13}\,R_{12}$. Below, each complex angle will be decomposed into its real and imaginary parts, namely $z_{ij}=z_{ij_R} + z_{ij_I}\, i$. For our numerical study we will calculate the Yukawa couplings from this parametrization, taking for $U$ and the neutrino mass-squared differences the expression and central values given in~\cite{ParticleDataGroup:2024cfk}. The only free parameters remaining in $U$ will be the two Majorana phases, $\eta_{1,2}$.

As other models for neutrino masses, the scotogenic model also has the virtue of incorporating a baryogenesis mechanism~\cite{Fukugita:1986hr}. Namely, as the sterile neutrinos decay in the primitive Universe, generally some amount of lepton asymmetry is produced, which is partially converted into a baryon asymmetry by the electroweak sphaleron processes~\cite{Kuzmin:1985mm}. The baryon asymmetry freezes at a temperature $T \sim 130$~GeV~\cite{DOnofrio:2014rug}, when the Hubble rate becomes faster than the sphaleron processes. 

Baryogenesis can be described by a set of BEs. 
At the low temperatures we are interested, all charged lepton Yukawa interactions are fast and the evolution of the lepton asymmetry can be followed appropriately by one BE for each lepton flavor (see the references on flavor quoted above). It also becomes necessary to trace the sterile neutrino abundances and, finally, we will add a BE to track the evolution of the asymmetry in the inert Higgs field, which plays a key role as we will show. Under several -reasonable and common- approximations, including kinetic equilibrium, Maxwell-Boltzmann statistics, neglect of several scattering processes and working at linear order in the asymmetries (see e.g.~\cite{Kolb:1979qa, Luty:1992un,  Giudice:2003jh,Buchmuller:2004nz,Davidson:2008bu}), 
an appropriate set of BEs for our study is 
\begin{eqnarray}
\label{eq:boleq}
\frac{\dif Y_{N_i}}{\dif z} &=&\frac{-1}{sHz}
\left(\frac{Y_{N_i}}{Y_{N_i}^{eq}}-1\right) \sum_\alpha 
2 \g{N_i}{\ell_\alpha \eta} \;, \nonumber \allowdisplaybreaks \\
\frac{\dif Y_{\Delta_\alpha}}{\dif z} &=& \frac{-1}{sHz} \left\{
\sum_i \left( \frac{Y_{N_i}}{Y_{N_i}^{eq}} - 1
\right)\epsilon_{\alpha i} \, 2 \g{N_i}{\ell_\alpha \eta} \right. \nonumber \\ 
 &-&\left.   \sum_i \g{N_i}{\ell_\alpha \eta} \left[y_{\ell_\alpha} + y_{\eta} \right] \right. \nonumber\\ 
&-& \left.   \sum_{\beta} \left( 
\gp{\ell_\alpha \eta}{\bar \ell_\beta \bar \eta} + (1+\delta_{\alpha \beta}) \,
\g{\ell_\alpha \ell_\beta}{\bar \eta \bar \eta}  
\right) [y_{\ell_\alpha} + y_{\ell_\beta} + 2 y_{\eta}]
\right. \nonumber\\ 
&-& \left.   \sum_{\beta \neq
\alpha} \left( 
\gp{\ell_\alpha \eta}{\ell_\beta \eta} + 
\g{\ell_\alpha \bar{\eta}}{\ell_\beta \bar{\eta}} + 
\g{\ell_\alpha \bar{\ell_\beta}}{\eta\bar{\eta}} 
\right) [y_{\ell_\alpha} - y_{\ell_\beta}] \right\} \; , \nonumber \allowdisplaybreaks \\
\frac{\dif Y_{\Delta \eta}}{\dif z} &=& \frac{1}{sHz} 
\Biggl\{
\sum_{i,\alpha} \left( \frac{Y_{N_i}}{Y_{N_i}^{eq}} - 1
\right)\epsilon_{\alpha i} \, 2 \g{N_i}{\ell_\alpha \eta} \Biggr. \nonumber \\ 
&-& \left.   \sum_{i,\alpha} \g{N_i}{\ell_\alpha \eta} \left[y_{\ell_\alpha} + y_{\eta} \right] \right. \nonumber\\ 
&-& \left.   \sum_{\alpha,\beta}\left( 
\gp{\ell_\alpha \eta}{\bar \ell_\beta \bar \eta} + 
2\, \g{\ell_\beta \ell_\alpha}{\bar \eta \bar \eta}  
\right) [y_{\ell_\alpha} + y_{\ell_\beta} + 2 y_{\eta}]
\right. \nonumber\\ 
&-& \Biggl.  2 \left( 
\g{\bar \phi \eta}{\phi \bar \eta} + 
2 \, \g{\eta \eta}{\phi \phi} 
\right) [y_{\eta} - y_{\phi}] \Biggr\} \; ,
\end{eqnarray} 
with $i=1,2,3$, $\alpha=e, \mu, \tau$, and where $z \equiv M_1/T$, $Y_X \equiv n_X/s$ is the number density of  particle $X$ normalized to the entropy density, and $y_X \equiv Y_{\Delta_X}/Y_X^{eq} \equiv (Y_X - Y_{\bar X})/Y_X^{eq}$ is the corresponding number density asymmetry normalized to the equilibrium density. For $X \neq N_i$ we use the convention that $Y_X$ gives the density of a single degree of freedom of $X$.
We have also defined $Y_{\Delta_\alpha} \equiv Y_B/3 - Y_{L_\alpha}$, where $Y_B$ is the baryon asymmetry defined previously and $Y_{L_\alpha}=(2y_{\ell_\alpha}+y_{e_{R\alpha}})Y^{eq}$ is the total lepton asymmetry in flavor $\alpha$ normalized to the entropy density (with $Y^{eq} \equiv Y_{\ell_\alpha}^{eq} = Y_{e_{R\alpha}}^{eq}$ and $e_{R\alpha}$ the right-handed charged leptons). Finally, we have also introduced the notation $\g{a, b, \dots}{c, d, \dots} = \gamma(\proname{a, b, \dots}{c, d, \dots})$ for the reaction densities (i.e.~rates per unit volume) of the corresponding process. The prime on some rates indicates that the contribution from on-shell neutrinos has to be subtracted~\cite{Kolb:1979qa, Giudice:2003jh}, and the term with the Kronecker delta ($\delta_{\alpha \beta}$) takes into account that processes with $\alpha=\beta$ change $Y_{\Delta_\alpha}$ by two units. 

We have calculated the cross sections analytically and then integrated them numerically to obtain the reaction densities (following mainly~\cite{Giudice:2003jh}, but neglecting finite temperature effects). The processes mediated by the sterile neutrinos, like $\proname{\ell_\alpha \eta}{\bar \ell_\beta \bar \eta}$, have contributions form $N_2$ and $N_3$ (the contribution from $N_1$ is negligible for the tiny values of $N_1$-Yukawa couplings to be considered). Therefore the reaction densities can be written as the sum of three contribution, e.g.~$\g{\ell_\alpha \eta}{\bar \ell_\beta \bar \eta}={\g{\ell_\alpha \eta}{\bar \ell_\beta \bar \eta}}(N_2)+\g{\ell_\alpha \eta}{\bar \ell_\beta \bar \eta}(N_3)+\g{\ell_\alpha \eta}{\bar \ell_\beta \bar \eta}(N_2,N_3)$,
where the last term comes from the interference of the amplitudes mediated by $N_2$ and $N_3$. For solving the BEs we have approximated this interference part by $\g{\ell_\alpha \eta}{\bar \ell_\beta \bar \eta}(N_2,N_3) \approx 2 \, \cos{\xi} \, \sqrt{\g{\ell_\alpha \eta}{\bar \ell_\beta \bar \eta}(N_2)} \, \sqrt{\g{\ell_\alpha \eta}{\bar \ell_\beta \bar \eta}(N_3)}$, where $\xi$ is the phase of the interference term. This approximation works well for the benchmark points below, because the lepton asymmetry is ultimately generated at $T \ll M_{2,3}$ and in that limit the approximation becomes exact, since the dependence on $M_2$ and $M_3$ factors out of the integrals (then the rates tend to expressions similar to the approximations given e.g.~in~\cite{Buchmuller:2004nz, Hugle:2018qbw}, but  with a Boltzmann suppression for $T \lesssim m_\eta$). It is also worth noticing that for higher temperatures, where this approximation might become inaccurate, the dominant washouts actually come from inverse decays of $N_2$ and $N_3$ (which have been included in our set of BEs). 

The CP asymmetries in $N_i$ decays, $\epsilon_{\alpha i}$, are defined as
\begin{equation*}
\epsilon_{\alpha i} \equiv 
\frac{\displaystyle \Gamma(\proname{N_i}{\ell_\alpha \eta})
- \Gamma(\proname{N_i}{\bar \ell_\alpha \bar \eta})}
{\displaystyle \sum_\beta
    \Gamma(\proname{N_i}{\ell_\beta \eta} )+ \Gamma(\proname{N_i}{\bar
    \ell_\beta \bar \eta})} \; ,
\end{equation*}
where $\Gamma(\proname{N_i}{x\, y})$ denotes the decay rate of $N_i$ into $x\, y$.
From the one-loop contributions we get, assuming that $(M_j - M_i)$ is much larger than the total decay widths,
\begin{eqnarray*}
\epsilon_{\alpha i}&=&
\sum\limits_{j \neq i} \frac{\miim{h_{\alpha j} h_{\alpha i}^* (h^\dag h)_{i j}}}{8 \pi (h^\dag h)_{ii}} \left[ \frac{M_i M_j}{M_j^2 - M_i^2} \left(1-\frac{\mhdos^2}{M_i^2}\right)^2  \right. \nonumber \\  \Biggl. && \qquad -\frac{M_j}{M_i} f(\tilde M_i^2/M_i^2,\tilde M_j^2/M_i^2)\, \Biggr]  \nonumber \\ &+&\sum\limits_{j \neq i} \frac{\miim{h_{\alpha j} h_{\alpha i}^* (h^\dag h)_{ji}}}{8 \pi (h^\dag h)_{ii}} \frac{M_i^2}{M_j^2 - M_i^2} \left(1-\frac{\mhdos^2}{M_i^2}\right)^2 \,,
\label{eq:epsi}
\end{eqnarray*}
 where $f(x,y)=1+(1/x+y/x^2)\ln{[1-x^2/(x+y)]}$ and $\tilde M_k^2 \equiv M_k^2 - \mhdos^2 \, (k=1,2,3)$. Note that for $\mhdos = 0$ this expression is equal to the obtained in~\cite{Covi:1996wh}) and that the last term is purely flavored, i.e.~it is zero when summed over all lepton flavors.

The set of BEs can be closed by additional relations among the number density asymmetries, which are obtained considering fast interactions 
(that lead to equalities among chemical potentials) and the conserved charges~\cite{Harvey:1990qw}. Performing an analysis quite similar to the lowest temperature regime described in~\cite{Nardi:2006fx}, but including the additional scalar doublet $\eta$, we obtain  
\begin{eqnarray}
Y_{\ell_\alpha} &=& \frac{32(Y_{B-L}-Y_{\Delta_\alpha})-442 Y_{\Delta_\alpha} - 27 Y_{\Delta \eta} }{1422}\;, \nonumber \\
Y_{\Delta \phi} &=& - \frac{8 Y_{B-L} + 13 Y_{\Delta \eta}}{79}\;,
\label{eq:chemical}
\end{eqnarray}
and
\begin{equation}
    Y_{B} =\frac{28 Y_{B-L} + 6 Y_{\Delta \eta}}{79}\;.
\end{equation}

\section{Spectator processes and benchmark points} 
\begin{table*}
\caption{\label{tab:bp}Benchmark points for leptogenesis in the scotogenic model. The Yukawa couplings have been obtained from the Casas-Ibarra parametrization given in Eq.~\eqref{eq:ci}, with $z_{12_R}=z_{12_I}=z_{23_R}=0$ for all points. Neutrino mixing data has been taken equal to the central values reported in the 2024 edition of the review of particle physics by the Particle Data Group~\cite{ParticleDataGroup:2024cfk} assuming a normal hierarchy. To evaluate $\Lambda$ we have taken $m_{\eta_I}^2 \simeq m_{\eta}^2$ and $m_{\eta_R}^2=m_{\eta_I}^2 + 2 \lambda_5 v^2$ in eq.~\eqref{eq:lambda}. Sphalerons freeze out at $T=130$~GeV and for all points $Y_B\simeq8.7 \times 10^{-11}$.}
\begin{ruledtabular}
\begin{tabular}{cccccccccccc}
 BP&$M_1$[TeV]&$M_2$[TeV]&$M_3$[TeV]&$\mhdos$[TeV]&$\lambda_5$&$z_{13_R}$&$z_{13_I}$&$z_{23_I}$&$\eta_1$&$\eta_2$&$m_1$[eV]\\ 
 \hline
1&1.3&6&12&0.6&$1.4\times10^{-5}$&$9.5\times10^{-6}$&$9.5\times10^{-6}$&0&$\pi/2$&0&$10^{-15}$\\
2\footnote{CR($\mu-e$, Ti)=$1.3\times10^{-19}$.}&3&6&9&0.9&$1\times10^{-5}$&$3\times10^{-6}$&$-3\times10^{-6}$&2.5&1.2&0&$10^{-13}$\\
3\footnote{Br($\mu \to e \gamma$)=$5\times10^{-13}$. CR($\mu-e$, Ti)=$2.1\times10^{-15}$, Br($\mu \to 3e$)=$3\times10^{-15}$ (dipole contributions).}&4&5.5&8&1.6&$1.23\times10^{-4}$&$6.5\times 10^{-6}$&$-6.5\times10^{-6}$&6&$\pi/2$&0&$10^{-12}$\\
4\footnote{Zero initial abundance of $N_1$, CR($\mu-e$, Ti)=$9.3\times10^{-20}$.}&2.5&$\sqrt{10} M_1$&$10 M_1$&$0.7 M_1$&$1\times10^{-3}$&$6\times10^{-5}$&$-6\times10^{-5}$&5&$\pi/2$&$\pi/4$&$10^{-11}$\\
\end{tabular}
\end{ruledtabular}
\end{table*}

\begin{figure}[t]
\includegraphics[width=0.5\textwidth,{angle=0}]{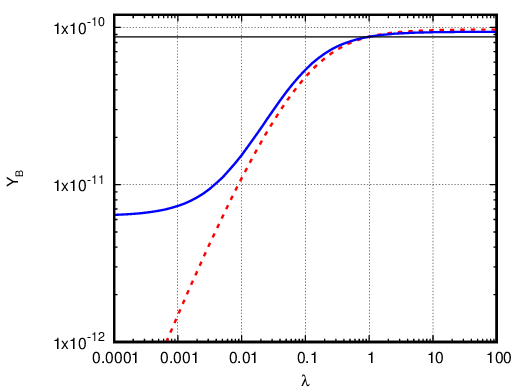}
\caption{The baryon asymmetry obtained for BP1 (solid blue curve) and BP4 (dashed red curve) integrating the BEs~\ref{eq:boleq}, but with the spectator processes reaction densities $\g{\bar \phi \eta}{\phi \bar \eta}$ and
$\g{\eta \eta}{\phi \phi}$ multiplied by $\lambda$, as a function of this parameter $\lambda$. For reference, the observed value of $Y_B$ is depicted with the horizontal solid black line.}
\label{fig:spect1}
\end{figure}

We have made a rough exploration of the parameter space compatible with the data on neutrino mixing, successful leptogenesis and a scalar DM candidate with 
$m_\eta \gtrsim 530$~GeV. Below we describe four benchmark points (BPs) that uncover interesting regions of parameter space and discuss the role of some spectator processes. The values of the parameters are given in Table~\ref{tab:bp}. 

BP1: This BP shows that leptogenesis in the scotogenic model is possible for masses 
as low as $M_1 \sim 1.3$~TeV, assuming a thermal initial abundance of $N_1$, i.e.~almost one order of magnitude below the lower bound found in~\cite{Hugle:2018qbw}. We stress that lowering the mass scale of (non-resonant) leptogenesis one order of magnitude, particularly at these already low energy scales for thermal leptogenesis, is not trivial. In the scotogenic model this can be achieved thanks to the sizable mass of one of the states coupled to the sterile neutrinos, namely the inert Higgs $\eta$, 
so that the washouts mediated by the sterile neutrinos are Boltzmann suppressed for $T \lesssim m_\eta$.  
This way of suppressing washouts can be very effective, but for a proper treatment it is crucial to consider the number density asymmetry of $\eta$ particles that is also generated during decays. This is because an excess number of, e.g., $\eta$ over $\bar \eta$, also induces washout of the lepton asymmetry, as it increases the rate of, e.g., $\proname{\ell_\alpha \eta}{\bar \ell_\beta \bar \eta}$ over $\proname{\bar \ell_\alpha \bar \eta}{\ell_\beta \eta}$.
The evolution of this asymmetry ($Y_{\Delta \eta}$) is regulated by the rates of the spectator processes $\proname{\bar \phi \eta}{\phi \bar \eta}$ and $\proname{\eta \eta}{\phi \phi}$, which are proportional to $\lambda_5^2$ (see the last  equation of the set~\eqref{eq:boleq}). If fast enough, they lead to an exponential depletion of the $\eta$-asymmetry, but if not, $Y_{\Delta \eta}$ is proportional to the lepton asymmetries. In the latter case, washout terms in the BEs like $\gp{\ell_\alpha \eta}{\bar \ell_\beta \bar \eta} y_\eta$ are not exponentially suppressed (recall that $y_\eta = Y_{\Delta \eta}/Y_\eta^{\rm eq}$).

To show the importance of these spectator processes, we have solved the BEs~\eqref{eq:boleq} for BP1, but multiplying the rates of $\proname{\bar \phi \eta}{\phi \bar \eta}$ and $\proname{\eta \eta}{\phi \phi}$ by an arbitrary parameter $\lambda$. The resultant baryon asymmetry as a function of $\lambda$ is represented by the blue curve in Fig.~\ref{fig:spect1}. In the limit $\lambda \to \infty$ spectator processes are faster than the Hubble rate during the whole leptogenesis epoch, ensuring the equality among the chemical potentials of the inert and SM Higgs fields, $\mu_{\eta}=\mu_{\phi}$, 
which implies a Boltzmann suppression of $Y_{\Delta \eta}$ relative to $Y_{\Delta \phi}$ for $T \lesssim m_\eta$. Instead, for $\lambda \to 0$ these spectator processes are absent, which effectively corresponds to the conservation of a modified lepton number, implying that $Y_{\Delta \eta}$ must be equal to a linear combination of $Y_{\Delta_{e,\mu,\tau}}$ and the corresponding washouts are not exponentially suppressed. 
For this BP there is a difference of more than one order of magnitude between the two limits and 
the actual asymmetry, i.e.~the one with $\lambda=1$, is close to the maximum value obtained when $\mu_\eta=\mu_\phi$. We note that this effect was discussed only qualitatively in~\cite{Racker:2013lua}
(which assumed the equality $\mu_{\eta}=\mu_{\phi}$). It was also presented more quantitatively in a proceedings~\cite{Racker:2014yfa}, but without any details on the BEs and in a toy scenario (not fitting neutrino data). 

\begin{figure}[t]
\includegraphics[width=0.5\textwidth,{angle=0}]{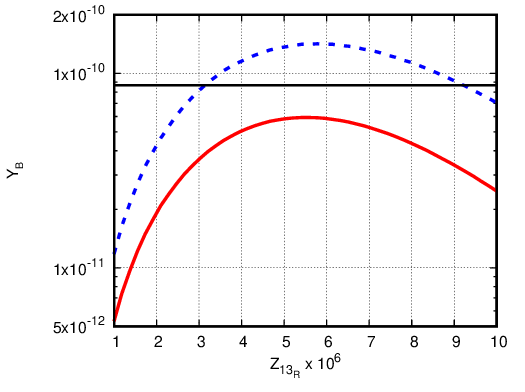}
\caption{The baryon asymmetry as a function of $z_{13_R}$ assuming $\mu_\phi=\mu_\eta$ (blue-dashed curve) and using the complete set of BEs~\ref{eq:boleq} (red-solid curve). Here we have taken $M_1= 1.6$~TeV, $M_2=\sqrt{10} M_1$, $M_3=10 M_1$, $m_{\eta}=0.6 M_1$, $\lambda_5= 8 \times 10^{-6}$, $\eta_1=\pi/2$, $\eta_2=0$, $m_1=10^{-13}$~eV, $z_{13_I}=z_{13_R}$ and $z_{12}=z_{23}=0$. Note that $z_{13_R} \in [10^{-6},10^{-5}]$. The observed value of $Y_B$ is depicted with the horizontal solid black line.}
\label{fig:spect2}
\end{figure}
In some regions of parameter space, the simplifying assumption $\mu_\eta=\mu_\phi$ (that allows to reduce the set of BEs) can lead to a considerable overestimate of the baryon asymmetry. This is illustrated in Fig.~\ref{fig:spect2}, where the baryon asymmetry obtained with this assumption is plotted as a function of $z_{13_R}$ and compared to  the more precise calculation using the full set of BEs~\eqref{eq:boleq}. 
For these points of parameter space $Y_B$ is overestimated by a factor of 2 to 3, giving in particular -wrong- values above the observed BAU for $3\times 10^{-6} < z_{13_R} < 9 \times 10^{-6}$.

BP2: The first BP is interesting because it shows that -and how- leptogenesis is possible with very low values of the sterile neutrino masses in one of the simplest extensions of the SM with a natural DM candidate.
However, the Yukawa couplings of all neutrino species are too tiny for them to leave signals in current or planned experiments. BP2 reveals that successful leptogenesis is possible with higher values of the Yukawa couplings of $N_2$ and $N_3$, large enough to induce CLFV processes that could be observed in the foreseeable future. More specifically, for this BP we obtain that the muon-to-electron conversion rate in titanium is larger than $10^{-19}$ (relative to the muon capture rate), above the sensitivity that could be achieved in planned experiments~\cite{CGroup:2022tli}. We have calculated the conversion rate via the relation
\begin{equation}
\label{eq:mueti}
{\rm CR}(\mu-e, {\rm Ti})={\rm Br}(\mu \to e \gamma) B(A,Z)/428,
\end{equation}
with $B(A,Z)=1.8$~\cite{Kuno:1999jp} (the lowest value quoted in that work). In turn, the branching ratio Br($\mu \to e \gamma$) has been computed using the analytical expression of~\cite{Toma:2013zsa}, taking $m_{\eta^\pm}^2 \simeq m_\eta^2$.

For this BP the imaginary part of the complex angle $z_{23}$ is moderately large ($z_{23_I}=2.5$), which results in Yukawa couplings larger by a factor $\sim \cosh{z_{23_I}}$ than a rough estimate from Eq.~\eqref{eq:mnu} disregarding complex phases. Therefore, for experimental probes the regions of parameter space with large imaginary parts of complex angles are the most interesting ones. They may be motivated by symmetries, but also renormalization group running effects might be important (see~\cite{Merle:2015ica} and~\cite{Drewes:2021nqr} for a different parametrization of the orthogonal matrix $R$ motivated by symmetry considerations). 
In any case, here we have shown that leptogenesis in the scotogenic model is possible for Yukawa couplings large enough to allow for experimental exploration and in regions of parameter space different from those allowed by resonant leptogenesis or leptogenesis via neutrino oscillations.

BP3: The Boltzmann suppression of washouts due to the mass of the inert Higgs can be very effective, enabling successful leptogenesis for Yukawa couplings much larger than those of BP2. 
For BP3 $z_{23_I}=6$ and the Br($\mu \to e \gamma$)=$5\times10^{-13}$, which is above the current bound, Br$(\mu \to e \gamma) < 4.2 \times 10^{-13}$ (90\% confidence level)~\cite{MEG:2016leq}. Therefore this BP shows that some regions of parameter space compatible with successful leptogenesis are already ruled out by experiments.  
The CR($\mu-e$, Ti) obtained from Eq.~\eqref{eq:mueti} is below the current bound, but much larger than sensitivities expected in planned experiments. 
Actually, this expression for the CR($\mu-e$, Ti) only takes into account the dipole contribution, but there is also a non-dipole contribution that is dominant in some regions of the parameter space of the scotogenic model~\cite{Toma:2013zsa}, leading to much larger values of the $\mu-e$ conversion rate than the obtained from Eq.\eqref{eq:mueti}. 
Given that successful leptogenesis is possible for such large Yukawa couplings, the possibility opens to explore regions of parameter space where the ratio CR($\mu-e$, Ti)/Br$(\mu \to e \gamma)$ departs from the relation given in Eq.~\eqref{eq:mueti} and becomes a more distinct signal of the scotogenic model. Similar considerations apply to the ratio Br($\mu \to 3e$)/Br$(\mu \to e \gamma)$, given that the Br($\mu \to 3e$) also is above the sensitivity of planned experiments for this BP (see~\cite{Toma:2013zsa}).

BP4: For the previous BPs we have assumed a thermal initial abundance of the singlet fermion $N_1$ (as was also assumed in~\cite{Hugle:2018qbw} and in many other models of low-scale baryogenesis from particle decays). This amounts to assume that the sterile neutrinos were produced at higher temperatures by some other -unknown- process not related to the Yukawa interactions of the scotogenic model.
With BP4 we show that, perhaps unexpectedly, low-scale leptogenesis with $M_1 \sim 2.5$~TeV is also possible starting with a zero initial abundance of $N_1$, i.e.~with neutrino production being due just to the Yukawa interactions of the model. This requires large values of $z_{23_I}$ and for this BP the CR($\mu-e$, Ti) lies within the limit of the sensitivity of planned experiments. The spectator processes are even more important here, as can be seen from the red curve in Fig.~\ref{fig:spect1}, which has been obtained in an analogous way to the blue curve for BP1. Note that, although $\lambda_5$ is two orders of magnitude larger than for BP1, the baryon asymmetry decreases by one order of magnitude for $\lambda \sim 0.01$ and by several orders of magnitude if the spectator processes are turned off. 

As a final comment, we note that for all the BPs above, the electric dipole moment of the electron estimated from~\cite{Abada:2018zra} is -at least- a few orders of magnitude below expected sensitivities and therefore we have not quoted it (e.g., for BP3 $d_e/e$ is below $5\times 10^{-33}$~cm). 

\section{Conclusions}
To conclude, we have shown that successful leptogenesis in the scotogenic model is possible for masses as low as 1.3 (2.5)~TeV for thermal (zero) initial abundance of the lightest sterile neutrino, and for Yukawa couplings large enough to induce CLFV processes in current and planned experiments. The regions of parameter space we have uncovered -which are compatible with neutrino masses and scalar DM- do not involve quasi-degenerate neutrinos. Instead, washouts are suppressed exponentially as the result of the sizable mass of the inert Higgs and the action of some spectator processes. The treatment presented here for this suppression mechanism, which involves a massive particle different from its antiparticle, may apply to baryogenesis in other extensions of the SM.

\begin{acknowledgments}
J.R. thanks the support by a grant Consolidar-2023-2027 from the Secretaría de Ciencia y Tecnología (SeCyT) de la Universidad Nacional de Córdoba (UNC) and also support from the Consejo Nacional de Investigaciones Científicas y Técnicas (CONICET), Argentina.
\end{acknowledgments}

\bibliography{lepto_ID}

\end{document}